\documentclass[epj,nopacs]{svjour}
\usepackage{graphics,epsfig,amsmath,amssymb,color}
\begin{document}

%
 % A useful Journal macros
 %%\def\jour#1#2#3#4{{#1} {\bf#2}, #4 (19#3)}
\def\jour#1#2#3#4{{#1} {\bf#2} (19#3) #4}
\def\jou2#1#2#3#4{{#1} {\bf#2} (20#3) #4}
 %Some Journal names used
\def\AP{Acta Phys. Pol. {B}}
\def\CP{Comp. Phys. Comm.}
\def\EPJ{Eur. Phys. J. {C}}
\def\EPJA{Eur. Phys. J. {A}}
\def\IJ{Int. J. Mod. Phys. {A}}
\def\IAN{Izv. Akad. Nauk: Ser. Fiz.}  
\def\JL{JETP Lett.}
\def\JP{J. Phys. {G}}
\def\ML{Mod. Phys. Lett. {A}}
\def\NC{Nuovo Cim. {A}}
\def\NIM{Nucl. Instr. Meth. {A}}
\def\NPA{Nucl. Phys. {A}}
\def\NP{Nucl. Phys. {B}}
\def\NPBP{Nucl. Phys. {B} (Proc. Suppl.)}
\def\NCS{Nuovo Cimento,  Suppl.}
\def\NPYF{Sov. J. Nucl. Phys.}
\def\Phy{Physica {A}}
\def\PL{Phys. Lett.  {B}}
\def\PR{Phys. Rev.}
\def\PRC{Phys. Rev. {C}}
\def\PRD{Phys. Rev. {D}}
\def\PRL{Phys. Rev. Lett.}
\def\PRp{Phys. Rep.}
\def\PPNP{Prog. Part. Nucl. Phys.}
\def\RPP{Rep. Prog. Phys.}
\def\Usp{Physics - Uspekhi}
\def\ZP{Z. Phys.  {C}}
%
% definitions
%
\def\nopar{\noindent}
\newcommand{\bec}{Bose-Einstein Correlations}
\def\Ecm{\sqrt{s}}
\def\snn{\sqrt{s_{\rm NN}}}
\def\spp{\sqrt{s_{\rm pp}}}
\def\see{\sqrt{s_{\rm ee}}}
\def\ep{$\rm{e}^+\rm{e}^-$}
\def\app{${\bar {\rm p}}{\rm p}$}
\def\vs{\vspace*}
\def\hs{\hspace*}
\def\bi{\bibitem}
\def\fig{Fig}
\def\sect{Sect.}
\def\col{Collab.}
\def\ea{{\sl et al.}}
\def\eg{{\sl e.g.}}
\def\vrs{{\sl vs.}}
\def\ie{{\sl i.e.}}
\def\va{{\sl via}}
\def\ct{\cite}
\def\lss{\pm\,\pm}
\def\al{\langle}
\def\ar{\rangle}
\def\leq{\leqslant}   
\def\geq{\geqslant}
\def\lssim{\lesssim}
\def\gtsim{\gtrsim}
\def\pythia{{\sc Pythia}}
\definecolor{Grey}{gray}{0.6}

\title{Relating multihadron production in hadronic and nuclear collisions}
\author{
Edward K.G. Sarkisyan\inst{1,2,}\thanks{\email{sedward@mail.cern.ch}} 
\and 
Alexander S. 
Sakharov\inst{3,4,}\thanks{\email{Alexandre.Sakharov@cern.ch}}% etc
}                     % Do not remove
\institute{The University of Texas at Arlington, Department of Physics, 
 %Box 19059, 
Arlington, TX 76019, USA
 \and Department of Physics, CERN, CH-1211 Geneva 23, Switzerland
 \and Department of Physics, Wayne State University, Detroit, MI 48202, USA
 \and TH Division, Department of Physics, CERN, CH-1211 Geneva 23, 
Switzerland}
\date{Received: / Revised: }
\abstract{
The energy-dependence of charged particle mean 
multiplicity and pseudorapidity
density at midrapidity measured
 in nucleus-nucleus and (anti)proton-proton collisions are studied in the 
entire 
  %whole 
 available energy range. The study is performed using a  
model, which
considers the multiparticle production process according to the
dissipating energy of 
 the participants and their types, namely a 
 combination of the constituent
quark picture together with
  Landau relativistic
hydrodynamics. The model reveals interrelations between the 
variables under study measured in nucleus-nucleus and nucleon-nucleon 
collisions.
 Measurements in nuclear reactions are shown to be well reproduced by the
measurements in pp/\app\ interactions
  %common
  and the corresponding fits are presented.  Different observations in
other types of collisions are discussed in the framework of the proposed
model. Predictions are made for measurements at the forthcoming LHC
energies.
\PACS{
      {PACS-key}{discribing text of that key}   \and
      {PACS-key}{discribing text of that key}
     } % end of PACS codes
} %end of abstract
\maketitle
 {\bf 1. } Soft hadron multiparticle production is one of the most 
intriguing topics 
in
high-energy interaction studies. 
 Data have been
investigated in different types of interactions, ranging from
lepton-lepton to nucleus-nucleus interactions,  and over a large
energy span, covering several orders of magnitude.
 QCD, the theory of strong interactions, has provided 
partonic
description of many observations. However, the problem of soft 
multiparticle
production still 
 eludes a complete
 understanding and remains 
one of the   
challenging problems in high-energy
physics \ct{book}.
 The new high-energy data from LHC  provide an opportunity to look at 
the 
system under 
new conditions.
 Of special interest are nucleus-nucleus collisions,
probing nuclear matter at extreme 
conditions,
where new forms of matter are expected to be created 
at very high 
 densities and temperatures. The data available from RHIC 
experiments 
 allow an interesting comparison of
 the particle production mechanisms
 with the less complex
\ep\ and pp systems.  In this context, the global, 
or bulk, variables such as the 
average
charged particle multiplicity and particle densities (spectra), which 
are the first available
experimental observables, are 
of fundamental
 interest \ct{book,mult} 
 as they are 
sensitive to 
 the 
 underlying 
 interaction dynamics.

 In this paper we consider the  center-of-mass (c.m.) energy 
dependence of 
 the average multiplicity and near 
midrapidity 
density of charged hadrons produced in nucleus-nucleus and 
(anti)proton-proton  collisions. Whereas the multiplicity is 
 sensitive 
 mostly
 to the fraction of  
energy being transformed into observed particles
 in a given reaction,  
 the midrapidity 
density
 reflects different stages of the reaction. 
  Both 
variables increase with the collision c.m. energy.
 Recent measurements at RHIC 
 follow the trends observed in \ep\ and proton-proton interactions.
  The values of both
bulk variables 
are 
found \ct{ph-sim,ph-rev} to be similar when comparing the measurements 
in \ep\
 interactions at the c.m. energy of $\see$, and in most central 
(``head-on'') heavy-ion collisions at the nucleon-nucleon c.m.energy 
 $\snn=\see$, where the measurements in the latter case are normalized 
to the 
number of pairs of participants (``wounded'' nucleons \ct{woundN}). 
 This phenomenon is found to be independent of
the 
type of 
colliding nucleus  
 for 
 $\snn$  between $\sim 20$~GeV and 200~GeV.
 
Assuming a universal mechanism of hadron production is present in both 
types of 
interaction, and that it is 
 driven
 only 
 by 
 the amount of energy 
 involved 
 into 
 secondary production,
one would expect the same value of the observables to 
be 
obtained in proton-proton collisions 
 when 
 $\spp$ is almost equal to $\snn$.
 However, 
comparing these  data 
 \ct{ua5-53900,cdf} to
the 
 measurements
from RHIC, 
one 
 finds
\ct{ph-sim,ph-rev,ph-56130,phxAph,phx-syststuds,phx-milov,phx-rev,br-rev}
 significantly lower values in hadron-hadron collisions.
 Furthermore, the 
 recent RHIC data from deuteron-gold 
interactions at 
$\snn=$ 200~GeV unambiguously points to the same values of the mean 
multiplicity as measured in antiproton-proton collisions 
\ct{ph-rev,ph-dAu}.

To  interpret these findings, 
 we have proposed in \ct{my}
  a phenomenological description based on the energy dissipation 
by
colliding participants into the state formed during the
  {\it early 
 stage} 
 of the collision.
  Particle production is then driven by the amount of the
initial effective energy deposited in this early phase 
by the 
{\it 
relevant types of participants}.
 The experimental observations referred to above  have been shown to be 
well described
 by this model
 and further predictions have been made.  In this
paper, the new and higher-energy data are added and 
 analyzed.
  \bigskip

 {\bf 2. } In 
 our consideration, the whole process of a collision is treated as the
expansion and the subsequent 
break-up into particles 
 from
  an initial state, in which the
 total available energy is assumed to be concentrated in a small 
Lorentz-contracted volume.
 There are no restrictions due to the conservation of quantum numbers
 other than energy and momentum constraints, thus allowing a relation 
between 
 the amount of energy deposited in the collision zone and the features of 
bulk
variables in different reactions.
 This approach resembles the Landau phenomenological hydrodynamical
description of multiparticle production in relativistic particle
collisions \ct{landau}.
 Though the hydrodynamical description does not match ideally the data on
multiparticle production in the whole range of pseudorapidity and 
different particle species, 
  it   
 gives good agreement with the multiplicity measurements in such different
reactions as nucleus-nucleus, pp, \ep\ and $\nu$p collisions
 demonstrating striking predictive power 
\ct{ph-rev,phx-rev,feinberg1,landau-exp-carr,landau-exp-st,busza,br-200,ph-mult}.
 Recently, the Landau model prediction for the Gaussian 
pseudorapidity shape due to  
 the longitudinal particle 
transport has been shown to reproduce well 
 the RHIC data \ct{br-rev,br-meson} as well as 
\ct{br-200,ph-mult} 
  the phenomenon 
 known as the ``limiting fragmentation'' \ct{limfrag}. The latter was 
 demonstrated to be independent of the energy and 
types 
of
colliding objects \ct{ph-rev,landau-exp-st,busza}.
 This indicates that the main assertions of the Landau approach are useful 
to estimate fractions of the energy dissipated into particles produced in 
different reactions, particularly in nucleus-nucleus collisions 
\ct{bjorken}.  
 Let us stress here that in this paper the Landau hydrodynamical model is
considered in the frame of the constituent quark picture as it is 
described
below.

 Once the collision of the two Lorentz-contracted
particles has resulted in a  fully thermalized system, but 
before 
 expansion, we
  assume 
that the production of secondary particles is defined by the 
fraction
of energy of the participants deposited in the volume of thermalized 
system at 
the 
moment of collision. This implies that there is a difference between 
results of 
collisions of structureless 
 and
 composite particles:
in composite particle collisions
not all the constituents deposit their energy when they form a 
small
Lorentz-contracted volume of the thermalized initial state.
 Therefore, in nucleon-nucleon collisions the interactions occur between 
single constituent, or dressed, quarks
 in accordance with the additive quark picture 
\ct{constq}, 
 and the other
quarks are considered to be spectators.  Thus the energy of the initial 
thermalized 
state which is
responsible for the number of produced secondary particles is that 
 of the interacting single quark pair. 
 The quark spectators which are not part of 
 the thermalized volume at the moment of collision
 do not participate in secondary particle
production.
 As a result, the leading particles \ct{leadp} resulting from the 
spectator 
quarks 
 carry away a significant 
part of the energy.
 Thus, only about
 1/3 of the entire nucleon energy is available for 
particle
production in pp/\app\ collisions.

 In heavy ion collisions, however, more than one quark
per nucleon interacts due to the large size of the nucleus
and to the long travel path  inside the nucleus. The more central 
the 
nucleus-nucleus collision is, the more interactions occur and the larger 
is the 
energy 
 available for secondary particle production.
 In central nuclear collisions, a contribution of constituent quarks 
rather than participating nucleons
seems to determine particle production and their 
 distributions 
\ct{voloshin}.
In
the most central collisions, the density of matter is so 
high (almost saturated)  that all three constituent quarks 
from each nucleon may participate 
nearly simultaneously 
in the collision, 
depositing their
 energy coherently into the thermalized collision volume. 
   In this case,
 the entire energy
of the participating nucleons 
 is
 available for bulk production in head-on nucleus-nucleus collisions. 
 Comparing this to 
 proton-proton collisions,  where only one out of 
three 
constituent quarks from each proton
interacts, one  
expects the features of the bulk variables per pair of participants 
measured in the 
most central 
heavy-ion 
interactions 
 to be similar to those  
from
proton-proton collisions but at a three times larger c.m. energy, 
$\spp \simeq 
3\, \snn$.

Adding together the above discussed ingredients, 
 namely the Landau model and the constituent quark picture,
 one finds  for the ratio of the charged particle
rapidity density $\rho(y)=(2/N_{\rm part})dN_{\rm ch}/dy$ per participant 
pair 
at the 
midrapidity value $y=0$ 
in 
heavy-ion reaction, $\rho(0)$, to the density $\rho_{\rm pp}(0)$ 
in 
pp/\app\ interaction,
 \begin{equation}
\frac{\rho(0)}{\rho_{\rm pp}(0)}=
  \frac{2\,N_{\rm ch}}{N_{\rm part}\, N^{\rm pp}_{\rm ch}}
 \, 
\sqrt{\frac{L_{\rm pp}}{L_{\rm NN}}}\,. 
\label{rap0}
\end{equation}

\nopar
Here,  $N_{\rm part}$ is the number of participants 
 ($N_{\rm part}= 2$ in nucleon-nucleon interactions), 
$N_{\rm ch}$ and $N_{\rm ch}^{\rm pp}$ 
 are the mean 
multiplicities in nucleus-nucleus and pp/\app\ interactions, respectively, 
and $L= \ln \frac {\sqrt {s}}{2m}$ with $m$ being the mass of a
participant, \eg\ 
$m=m_{\rm p}$, the mass of the proton, 
in nucleus-nucleus 
collisions.  
 According to our model, we compute the ratio (\ref{rap0}) 
 for the 
rapidity 
density  $\rho(0)$
and the multiplicity $N_{\rm ch}$ 
at $\snn$  and the rapidity density 
  $\rho_{\rm pp}(0)$ and the multiplicity $N_{\rm ch}^{\rm pp}$
at $3\,\snn$. Due to the above, we consider a 
constituent quark of mass $\frac{1}{3}m_{\rm p}$ as a participant in 
pp/\app\ 
collisions, and a proton as an effectively structureless participant in  
head-on nucleus-nucleus 
collisions.
 Then, from Eq.~(\ref{rap0}) one  obtains:
 \begin{equation}
  \rho(0)= \rho_{\rm pp}(0) \,   
  \frac{2\,N_{\rm ch}}{N_{\rm part}\, N^{\rm pp}_{\rm ch}}
 \,
\sqrt{1-\frac{4 \ln 3}{\ln\, (4 m_{\rm p}^2/s_{\rm NN})} }\,, 
 \:\:\: \snn=\spp/3. 
\label{prap0}
\end{equation}

\nopar
 $ \rho(0)$ is thus calculated from the measured values of $\rho_{\rm 
pp}(0)$ 
and the 
multiplicities measured in both reactions.
 
 \begin{figure*}
 \hspace*{2.5cm}
\resizebox{0.71\textwidth}{!}{%
  \includegraphics{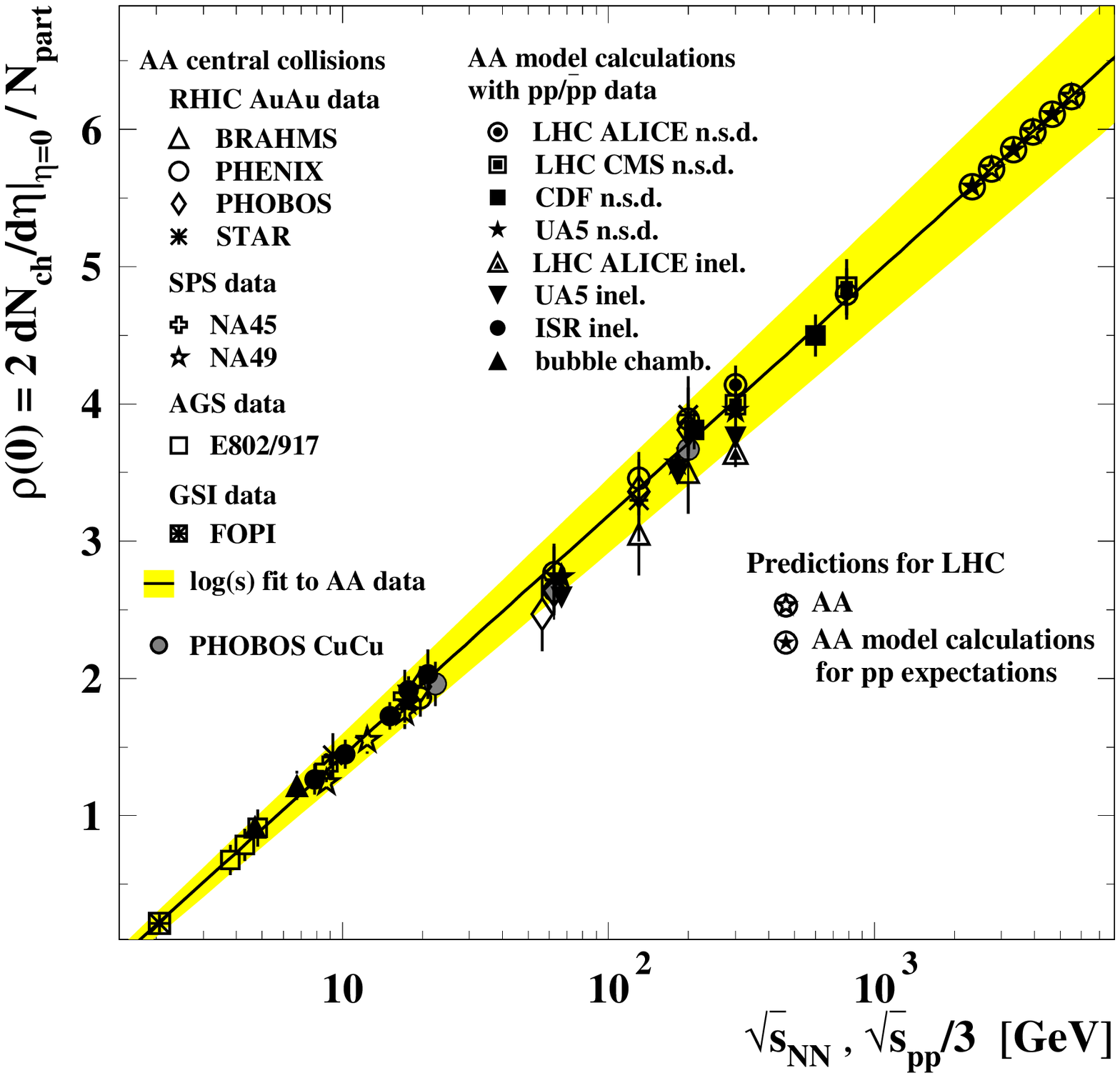} 
}
  \caption{   
 %\footnotesize
 \small
 The pseudorapidity density of charged particles per participant pair at
midrapidity as a function of c.m. energy per nucleon, $\snn$,  measured
in central nucleus-nucleus (AA) collisions and calculated from pp/\app\ 
data 
 using Eq. (\ref{prap0}).
 The AA data are from: 
 the AuAu measurements 
at
 RHIC by BRAHMS \ct{br-200}, 
 PHENIX \ct{phx-syststuds,phx-milov}, 
 PHOBOS \ct{ph-56130,ph-colgeom,ph-62,phobos-ncucu,phobos-etacucupT}, 
 and STAR \ct{star-200,star-pp-dAu-AuAu-3e,star9a2}
 experiments;   
 the values recalculated in \ct{phx-syststuds} from the measurements 
 at CERN SPS by CERES/NA45 \ct{na45} and 
 NA49 \ct{na49} experiments,
 at Fermilab AGS by E802 and E917 experiments \ct{ags}, 
 and at GSI by FOPI \col\ \ct{fopi}; 
 the PHOBOS data on CuCu collisions from 
\ct{phobos-ncucu,phobos-etacucupT,nouicer-rev}.
 The solid symbols show the 
 values obtained from Eq. (\ref{prap0}) using the following data on 
midrapidity 
densities measured in non-single diffractive collisions: pp data from 
ALICE \ct{alice900G2T} 
 and CMS \ct{cms900G2T} experiments at LHC and  from \app\ collisions by 
 UA5 \col\ at CERN SPS  \ct{ua5-53900,ua5-546} 
 and 
 ISR ($\spp=$~53~GeV),  
 by CDF \col\ at Fermilab \ct{cdf};
 and in the following inelastic collisions: \app\ data by UA5 
Collaboration 
 and  pp data from the LHC ALICE \ct{alice900G2T} 
 experiment and from the 
 ISR \ct{isr-thome}
 and bubble chamber \ct{fnal-rap205,fnalmult}
 experiments, 
   the latter 
 as recalculated in \ct{ua5-53900}.  
 The solid  line 
 shows the linear-log fit, $-0.33+0.38\ln(s_{\rm NN})$, to the AA data 
with
 the parameters and
 errors obtained 
  using a  combination of the data from the RHIC and SPS
 experiments.
 The shaded area shows  1-$\sigma$ error band to the fitted parameters. 
 The circled stars show the heavy-ion predictions for LHC  AA 
collisions (open stars) and from the expected LHC pp collisions  
 (solid stars), both calculated from the fit.
 }
\label{fig:rap0}
\end{figure*}
%%%%

 \begin{figure*}
\resizebox{0.82\textwidth}{!}{%
 \hspace*{3.5cm}   \includegraphics{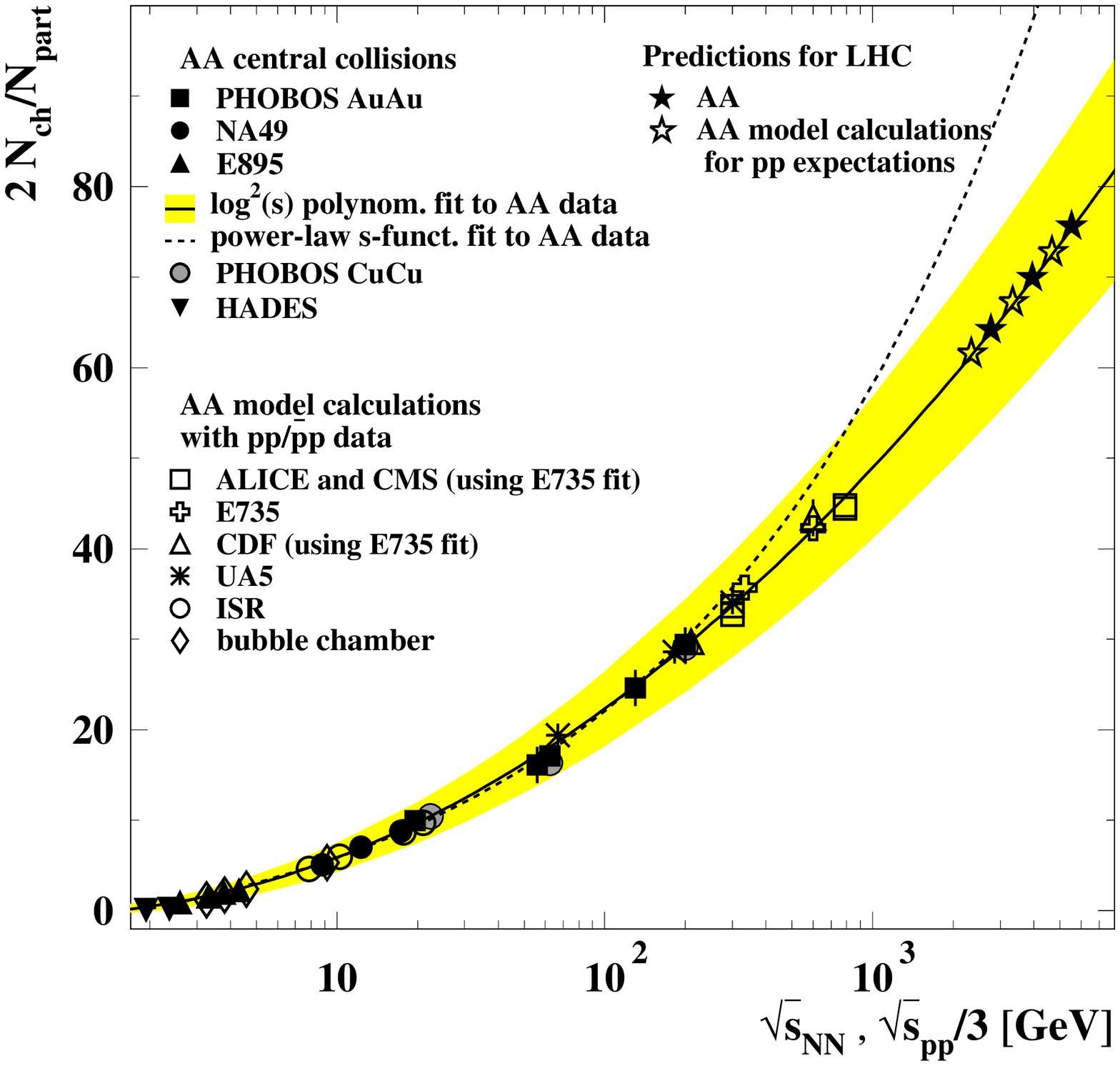}
}
\caption{ 
 \small
 The charged particle mean multiplicity per participant pair as a function
of the nucleon-nucleon c.m. energy, $\snn$, measured in 
the most central nucleus-nucleus (AA) collisions and calculated using 
pp/\app\ data
 from Eq. (\ref{pmult}).
 The solid symbols show the multiplicity values from the 
  AA data 
 as measured by PHOBOS \col\ at RHIC in AuAu  
 \ct{ph-sim,ph-rev,ph-mult,ph-colgeom,ph-62} and CuCu \ct{phobos-ncucu} 
 collisions, 
  by NA49 \col\ at CERN SPS \ct{na49-mult} 
 and by E895 \col\ at AGS \ct{agsmult} 
 (see also \ct{ph-sim}). The low-energy HADES measurements at GSI 
are 
taken from 
\ct{hades}. 
 The open symbols show the values obtained from  
Eq.~(\ref{pmult}) for the AA mean multiplicity based on: 
 \app\ collisions at FNAL by E735 \col\  
\ct{e735-conf-mult,pp-mult-rev},
 at CERN by  
 UA5 \col\ 
 at $\spp=$ 546 GeV \ct{ua5-546} 
 and $\spp=$ 200 and 900 GeV \ct{ua5-zp}; 
  pp collisions 
 at CERN-ISR 
   \ct{isr-thome}, and
 the ALICE,  CMS  and CDF multiplicities,  calculated using the 
E735 
 fit, $ 3.102\, s_{\rm pp}^{0.178}$ \ct{pp-mult-rev}, and the data 
 from bubble chamber experiments 
 \ct{fnalmult,bubblechamber}
the latter having been compiled and analysed
 in 
 \ct{eddi}. 
 The solid line shows the
 2nd-order log-polynomial fit $-0.35+0.24\ln(s_{\rm NN})+0.24\ln^2(s_{\rm 
 NN})$  to the AA data. 
 The shaded area shows 1-$\sigma$ error band to the fitted 
 parameters. 
 The dashed 
  line shows the power-law fit: $-7.32+5.92\,s_{\rm NN}^{0.174}$.
  The stars give the heavy-ion predictions for the LHC AA collisions 
(solid 
symbols) and those expected from 
 LHC pp collisions
 (open symbols)  based on the 2nd-order log-polynomial fit.
 }
\label{fig:multshe}
\end{figure*}
%%%

 Solving
Eq.~(\ref{rap0}) for $ N_{\rm ch}$ at given $\rho(0), \rho_{\rm 
pp}(0)$ and $N^{\rm pp}_{\rm ch}$ one  finds: 

 \begin{equation}
 \frac{2\, N_{\rm ch}}{N_{\rm part}} = 
N^{\rm pp}_{\rm ch} \,
  \frac{\rho(0)}{\rho_{\rm pp}(0)} \,   
\sqrt{1-\frac{2 \ln 3}{\ln\, (4.5 \snn/m_{\rm p})} }\,. 
\label{pmult}
\end{equation}
 \nopar 
 From these one calculates the expected values for nucleus-nucleus central
collisions,
 since they are related to pp/\app\ interactions through
 the constituent quarks picture of interactions 
 combined with
  the Landau
energy dissipation hydrodynamics.
 Note that the different sets of  observables used in
Eqs.~(\ref{prap0}) and (\ref{pmult}),  do
 not guarantee that both equations describe the data well, even if one of 
them does.
  \bigskip    

 {\bf 3. } 
  In this section, we consider
 the available data on the mean 
multiplicity and midrapidity density 
 measured in 
 nucleus-nucleus collisions. 

 Figure~\ref{fig:rap0} shows 
 the nuclear 
 data on 
 the pseudorapidity 
density, 
$\rho(0)$, 
per participant pair, 
 measured in head-on
nucleus-nucleus collisions (open symbols), 
 as  
 a function of  $\snn$ 
in the 
 energy range $\snn = 2 - 200$~GeV.
 The measurements are 
 from 
 experiments
 at 
GSI \ct{fopi},   
AGS \ct{ags},
CERN SPS 
 \ct{na45,na49}, 
 and RHIC
\ct{ph-56130,phx-syststuds,phx-milov,br-200,ph-colgeom,ph-62,phobos-ncucu,phobos-etacucupT,star-200,star-pp-dAu-AuAu-3e}.
 The data clearly satisfies a
linear-log relationship.
 Fitting, using a weighted combination of the data from the RHIC and SPS 
experiments, we find:
   $$\rho(0)=(-0.327 \pm 0.026)+(0.381\pm 0.021)\ln(s_{\rm NN}).$$
 Similar observations have been made in 
\ct{phx-syststuds,phx-milov,ph-62,nouicer-cu}. 
 Such a behaviour seems \ct{busza}
 to be well justified by the participant-driven
picture of the formation of the rapidity distribution and the $N_{\rm
part}$ scaling.
 In addition, as the nuclear data show, the process of bulk production
is largely characterized by $N_{\rm part}$, or soft scaling, rather than
by the number of binary collisions, $N_{\rm coll}$, 
 the latter being 
 attributed to hard scaling
\ct{phx-syststuds,ph-mult,ph-colgeom,ph-62,ph-iord,ph-softhard}.
 Recently, this feature has been confirmed by observing the $N_{\rm 
part}$ scaling in the ratio of the charged-particle yields measured  
at $\snn = 200$~GeV and 62.4~GeV, as   $\al p_T \ar$ 
increases 
from 0.25~GeV to $\sim 4$~GeV \ct{phobos-etacucupT}. The increase of the 
 ratio at large $\al p_T \ar$ 
 is believed 
to appear because of the harder spectra at higher $\snn$, so as to 
follow  
$N_{\rm coll}$ scaling and not to scale with $N_{\rm
part}$. Interestingly, the   $N_{\rm
part}$ scaling is observed to be the same for AuAu 
and 
CuCu interactions.

 Figure~\ref{fig:multshe} shows the $\snn$-dependence of the 
nucleus-nucleus data of the mean
multiplicity $N_{\rm ch}$ per participant pair from the most central
nucleus-nucleus collisions (solid symbols)  in the same $\snn$ range as
above.
 The data are taken from the measurements by the PHOBOS experiment at RHIC
\ct{ph-sim,ph-rev,ph-mult,ph-colgeom,ph-62,phobos-ncucu}, NA49 at CERN SPS 
\ct{na49-mult} and
by E895 at AGS \ct{agsmult}.
 The mean multiplicities are seen to increase as a 2nd-order 
logarithmic polynomial, and our fit gives:
$$N_{\rm ch}/{(0.5\,N_{\rm part}})=(-0.35 \pm 0.28) $$ $$ +\,\, (0.24 \pm 
0.16)\ln(s_{\rm NN})+(0.24 \pm 
0.03)\ln^2(s_{\rm NN}),$$ which well reproduces the 
energy-dependence of the measurements. This trend holds even for 
relatively 
low-energies, as demonstrated by the recent data by HADES 
{\ct{hades} displayed in 
Fig.~\ref{fig:multshe}.
 Such an $\ln^2(s_{\rm NN})$  behaviour is expected \ct{busza,unpart} to 
appear 
naturally 
as a
combination of the following features: the Landau model Gaussian 
shape
of the pseudorapidity distribution, the logarithmic increase of the
midrapidity density with the c.m. energy, and the limiting fragmentation.
 \bigskip

 {\bf 4. } 
 The  midrapidity density and the 
 multiplicity 
energy 
dependencies are analyzed in the framework of the model developed here, 
{\ie} applying 
Eqs. (\ref{prap0}) and (\ref{pmult}) to 
 calculate these variables in  nucleus-nucleus 
 interactions based on nucleon-nucleon data.

Using the pp/\app\ data on $\rho_{\rm pp}(0)$ from CERN
 \ct{ua5-53900,alice900G2T,cms900G2T,ua5-546,isr-thome}\footnote{
 For the LHC data  
  we exclude some results to avoid any dependence on the 
 experimental cuts
  and to have consistent data sets. We do not use     
  the 900 GeV data by ATLAS \ct{atlas900}  
  which applies a high
  $p_T$-threshold, and the inelastic (high-energy) data by ALICE
  \ct{alice7Tinel} where specific selection cut is used.
  This exclusion does not influence the consideration here.}
  and Fermilab
\ct{cdf,fnal-rap205,fnalmult}, the data on multiplicities 
$N_{\rm
ch}^{\rm pp}$ in pp/\app\ interactions 
\ct{ua5-546,isr-thome,fnalmult,ua5-zp,bubblechamber}, and 
$N_{\rm ch}$ 
from 
 Fig.  \ref{fig:multshe}, 
  along with the fact that the transformation factor from rapidity to
pseudorapidity does not influence the above ratio, we calculate, according
to Eq. (\ref{prap0}), the $\rho(0)$ 
 at $\snn=\spp/3$. 
 These calculations are shown by solid symbols in Fig.~\ref{fig:rap0}.
 One can see that the calculated $\rho(0)$ values
   are in a very good agreement with
  the  measured $\rho(0)$'s as well as with the obtained log-fit 
 in the whole measured  $\snn$ range.

 The agreement  is 
 more intriguing as one adds the hadronic data at c.m. energies higher
than the top RHIC energy, namely
 the densities for (anti)proton-proton interactions  
 at  
 $\spp=1.8$~TeV from Tevatron 
  \ct{cdf} and 
$\spp=2.36$~TeV
 \ct{alice900G2T,cms900G2T} 
 from the LHC. 
 In Fig.~\ref{fig:rap0} we compare the calculations using Eq. 
(\ref{prap0})
 on these TeV-energy data with our linear-log fit.
 One sees
the
 nucleus-nucleus $\rho(0)$'s 
 at $\snn=600$ and $\sim 800$~GeV,
 calculated from the highest energy hadron 
data, 
 agree well with the AA-fit. 
 This observation justifies the above conclusion from the  
lower-energy midrapidity density
dependence on the types of participants. 
 Note that Eq.~(\ref{prap0})  shows the 
 relevance
 of the Landau
hydrodynamical energy-dissipation $L$-factor which was required in order 
to correctly estimate
 the midrapidity density.

 Addressing now  Eq. (\ref{pmult}), we calculate the
participant-pair-normalized mean multiplicity $N_{\rm ch}/(0.5 N_{\rm
part})$ for nucleus-nucleus interactions from the pp/\app\ measurements 
\ct{ua5-53900,cdf,alice900G2T,cms900G2T,ua5-546,isr-thome,fnal-rap205,fnalmult}
of
$\rho_{\rm pp}(0)$
 and
     $N^{\rm pp}_{\rm ch}$,\footnote{
  For the ALICE, CMS and CDF multiplicities 
   the E735 power-law fit $N^{\rm pp}_{\rm ch}= 3.102\, 
 s_{\rm pp}^{0.178}$ \ct{pp-mult-rev} is used.}  
 and the
 corresponding $\rho(0)$ data 
 from Fig.  \ref{fig:rap0}.
  The calculated $N_{\rm ch}/(0.5 N_{\rm part})$ values 
  are shown in Fig.~\ref{fig:multshe} (open symbols) along with the
available nuclear data.
 One can see that the calculated $N_{\rm ch}/(0.5 N_{\rm
part})$ values
reproduce well the
$\log^2(s_{\rm NN})$ fit obtained here and follow the nuclear data 
points 
for $\snn=$
2~GeV to about 200~GeV.

The calculations for the new higher energy data
namely those for $\snn \gtsim $~200~GeV, 
are 
again of a special interest.
 These calculations help to check the proposed description over a larger 
energy range,
the possible fit and result in firmer predictions.
 Indeed, from Fig.~\ref{fig:multshe}, one can see that for $\snn \leq$ 200
GeV, it is quite difficult to distinguish between the two fits: the
power-law fit
 $
 \propto s_{\rm NN}^\gamma$, known to be a preferable fit to the 
 $N_{\rm pp}^{\rm ch}$ data \ct{ua5-546,pp-mult-rev}, 
 is almost as good as the 
 $\log^2(s_{\rm NN})$ 
 polynomial approximation.
  However, it is evident that, 
after inclusion of the new higher energy data,
 the $\log^2(s_{\rm NN})$ function is more preferable.
 \bigskip

 {\bf 5. }  
 From the above,  we conclude that using all the world available
measurements on the mean multiplicity and the midrapidity density in the
data of the 
 nucleon-nucleon and central nucleus-nucleus collisions, a 
clear
interrealtion between the two types of the data is obtained.
 This can be attributed to the universality of  the multiparticle
production process
 over almost three orders of magnitude
 of $\snn$. Under 
this assumption, predictions for the LHC energies can be made.

 Using the fits, shown in Figs.~\ref{fig:rap0} and ~\ref{fig:multshe}, 
and solving 
Eq.~(\ref{rap0}) for the midrapidity density $\rho_{\rm pp}(0)$ with 
$N_{\rm ch}^{\rm
pp}$ from the high-energy fit \ct{pp-mult-rev},
 the
expected $\rho_{\rm pp}(0)$ values for pp collisions at LHC are found to 
be 
about
5.8, 6.4, and 6.9 at $\spp= 7, 10$ and 14~TeV, respectively, within 
 5\% to 10\% uncertainties.   
 From the fit to the 
midrapidity densities $\rho(0)$, shown in Fig.~\ref{fig:rap0}, the 
$\rho(0)$ values, expected for 
PbPb collisions at LHC 
energies $\snn$, 
corresponding to the above $\spp$, are found to
be
about 5.7, 6.0, and 6.2 at $\snn=  2.76, 3.94$ and 5.52~TeV, 
respectively. 
The $\rho(0)$ predictions  are shown in 
Fig.~\ref{fig:rap0} by circled solid stars  
for PbPb interactions and 
by
circled open stars for those 
from the pp
expectations at LHC
at 
$\snn=\spp/3$ when calculated according  to our model.

 Comparing our predictions for  $\rho_{\rm pp}(0)$ to the predictions of 
other models and 
Monte 
Carlo tunes 
\ct{pp-mult-rev,atlas}, we find 
 that our values  are in the range 
of the midrapidity density values predicted
 there.
The values we find
 here
 are also consistent with 
 those from the CDF $\rho_{\rm pp}(0)$ fit \ct{cdf} 
 and from a similar, but higher-energy, CMS fit \ct{cms900G2T}.
  The values of $\rho(0)$ 
 we find 
for LHC heavy-ion 
collisions 
 are also well in the range of the expectations by different models
\ct{armesto-AAmult-rev,mitrovski}. 
 The $\rho(0)$ values obtained at $\snn\approx 5.5$~TeV  is
similar to that obtained by PHENIX from the fit to the  nuclear data
 \ct{phx-syststuds,phx-milov} and by PHOBOS from their extrapolation of
their AuAu data to PbPb collisions at LHC \ct{busza}. 

 Using the log$^2\,s_{\rm NN}$ fit to the mean multiplicity shown in 
Fig.~\ref{fig:multshe}, one finds the average multiplicity, 
$N_{\rm
ch}/(0.5 N_{\rm part})$, in PbPb collisions
to be about  64, 70 and 73 at $\snn= 2.76, 3.94$ and 5.52~TeV, 
respectively,
with 10\% to 15\% uncertainties. 
 The
$N_{\rm ch}^{\rm 
pp}$ 
in  pp collisions at LHC are expected to be about 73, 82 and 93
at $\spp= 7, 10$ and 14~TeV within about 10\% uncertainties, and are the 
same as one finds from 
the 
multiplicity high-energy power-law fit \ct{pp-mult-rev}.
 The $N_{\rm
ch}/(0.5 N_{\rm part})$ predictions
for PbPb collisions
are shown in 
Fig.~\ref{fig:multshe} by solid stars, while those expected from the 
LHC pp 
collisions at $\snn=\spp/3$ in the framework of our model,
are shown by open stars there. 

The $N_{\rm ch}/(0.5N_{\rm pp})$ we obtained 
are 
consistent with 
extrapolations from the present experimental measurements \ct{busza}.
 Our prediction at $\snn\simeq 5.5$~TeV is comparable
with the estimate from the pseudorapidity density spectra by PHENIX
\ct{phx-syststuds,phx-milov}. 
 Similar to the $\rho_{\rm pp}(0)$ values, the 
$N_{\rm ch}^{\rm pp}$  are within the range of the 
predictions by different models and 
Monte-Carlo tunes \ct{pp-mult-rev,atlas}. 
  \bigskip

 {\bf 6. } 
 Let us now dwell on some corollaries of the model proposed here and
discuss the results in view of other observations.

 From our consideration it follows that, at the same $\snn$, the 
mean
multiplicities as well as the midrapidity densities, normalized to the 
number of participants, would give the similar values when measured in 
central symmetric nucleus-nucleus collisions of different  
colliding nuclei, they are 
 largely driven by the initial 
energy  
deposited by the participants at early stage of collisions.
 Indeed, as seen from Figs.~\ref{fig:rap0} and \ref{fig:multshe},
this effect has been already observed at SPS energies
and now is confirmed by  the
  RHIC measurements 
at $\snn$ of about 50~GeV to 200~GeV. The same values for  
both the observables 
are obtained \ct{busza,nouicer-cu,nouicer-rev,nouicer-cu-npart} in Au-Au 
and Cu-Cu 
 data, as shown  in Figs.~\ref{fig:rap0} and \ref{fig:multshe}. 
Note that this effect has also been observed for the whole 
pseudorapidity region \ct{bb3}. Notice also the similarity in the 
above-discussed  
$N_{\rm part}$ dependence in low-$p_T$ \vrs\ high $p_T$ ranges  
\ct{phobos-etacucupT}.

  An interesting issue to be addressed in the framework of the 
 model is  to consider asymmetric collisions, such as
nucleon-nucleus 
 (pA/dA) ones.
 In such type of interactions, the bulk variables studied here, 
being measured at 
given 
$\snn$ are
expected to have the same 
values as those in pp/\app\ collisions at $\spp \simeq \snn$.
Indeed, 
assuming an   incident 
proton in p-nucleus collisions interacts in the same way it would interact 
in 
pp collision,  the
secondary particles in the reaction are assumed to be created 
 out  of the energy deposited by the interaction of a single pair of 
constituent 
quarks, one 
 from the proton and another one 
from a nucleon in the interacting nucleus. 
 This, in its turn,  
 implies  that the mean multiplicity and the midrapidity density 
  are  expected to be independent of the centrality of nucleus-induced  
collisions, $N_{\rm part}$ (within uncertainties due to intranuclear 
effects, \eg\ Fermi motion). 
   These expectation are shown 
to be well confirmed in the RHIC data on deuteron-gold interactions 
 at $\snn=$~200~GeV. 
 Moreover, the effect obtained at RHIC is shown 
 \ct{ph-rev,ph-dAu,phobos-ncucu} 
 to be  true also for hadron-nucleus collisions at lower $\snn\approx 
$~10--20~GeV. 

  In this study the Landau hydrodynamic model is used, leading to a good 
description of the data from different reactions
\ct{ph-rev,br-rev,feinberg1,landau-exp-carr,landau-exp-st,busza,br-200,ph-mult,br-meson}.
However, this is a 1+1 model and therefore does not 
take into account 
the transverse expansion of  the system which can be studied, for example, 
via 
the transverse energy, another important bulk observable.
Considering the 
measurements from SPS to RHIC of  
the  transverse energy midrapidity density, one  finds that  
this variable scales with the 
number of constituent quarks a way similar to that  of 
charged 
particle and photon midrapidity densities \ct{indet}. 
 Furthermore, 
as measured at RHIC, the 
ratio of 
the transverse energy  midrapidity density to that of multiplicity density 
is observed \ct{phx-syststuds} to be independent of  the number of 
participants, and  the transverse energy loss is found \ct{ph-RAA-AuCu} to 
be 
independent 
of the type of colliding nuclei if the same number of participating 
nucleons  is taken.   The observations indicate 
 scalings  
 of a similar nature to those considered here for multiplicities and 
midrapidity
densities. This seems also to reflect the fact that the inclusion of the
transverse expansion in the Landau model does not change the scaling of
the observables under study \ct{feinberg1,landau-pt}.  Currently, the 
model attracts
high interest and has undergone a generalization, see \eg\
\ct{landau-new}.

 As we have shown,  the constituent quark is a key component of 
 a correct
description of ``soft'' particle observables in particle and nuclear
collisions from a few GeV up to highest LHC energies, so that constituent
quarks 
 have to be taken as the interacting particles,  see \eg\
\ct{levin-bondarenko}. This already has  support from studies of the
multiplicity distributions in heavy-ion-induced interactions
\ct{voloshin,nouicer-cu-npart,bb3,bialasjpg,liu-universal}.  The
constituent quark picture has been exploited to reasonably model the
heavy-ion pseudorapidity and transverse energy data \ct{ind}. The elliptic
flow characteristics are also observed to scale when the constituent quark
frame is taken into account 
\ct{phobos-elipt-scal,steinb-qm09}.

The similarities of pp/\app\ and AA interactions
observed here for the two
basic variables suggest that the system is formed at early stage as a
superposition of contributions from the constituent quarks. The particle
multiplicities seem then to be derived by the total energy of 
participants,
available in the Lorentz-contracted volume. The importance of the {\it 
very early stage} of collision for soft particle production has been 
already discussed elsewhere 
\ct{bb3,bialasjpg,steinb-qm09,qcd-universality}. Due to the 
proportionality 
of the 
multiplicity to the entropy \ct{landau}, the multiplicity scaling, 
observed for 
different variables measured at RHIC, has been suggested to be connected 
to the 
total 
 produced entropy \ct{torrieri}.     

 Considering this and recalling the above-mentioned similarity in 
heavy-ion and \ep\ collisions observed, one would expect    
 the same model to be valid in matching the mean multiplicity and 
the 
midrapidity values in heavy-ion and \ep\ data. Indeed, as we have shown in 
\ct{my}, both variables follow the same energy dependence within the 
framework of our picture, as soon as  
one considers that the structureless electron and positron 
 deposit their total energy into the 
Lorentz-contracted volume
 similar to nucleons in central nuclear collisions. From this, the factor
1/3 applied to the pp/\app\ energy scale is expected to result in a 
good match between the \ep\ 
and pp/\app\ data on multiplicity and
midrapidity densities as shown in \ct{my}. This solves the
 problem with the energy-scaling factor of 1/2 used
 in \ct{ph-sim,ph-rev}, where the $\spp/2$ shift 
 is shown 
 to 
 provide a reasonable description of 
the average multiplicity c.m. 
 energy 
 dependence but 
not of 
the midrapidity 
density when comparing heavy-ion/\ep\ data to those from pp/\app\ 
collisions. 
  We recall that the energy-scaling factor 1/3, 
 has 
already been shown in
\ct{ee3pp,ee3pp-landau} to give 
 good agreement
 of the pp mean 
multiplicity
data relative to those from \ep\ annihilation, for a review see \ct{book}.
 It is remarkable that  
the 3NLO  perturbative QCD  \ct{dremin} fit to \ep\ data  
\ct{lep1.5-2adl}
  describes the pp/\app\ multiplicity data providing 
 the inelasticity is set to $\approx 0.35$ \ct{pp-mult-rev},  favouring 
the effective 1/3 
c.m. 
energy    in
  multihadron production  in pp/\app\ 
    reactions. 

 As already mentioned, the average multiplicity is defined mostly
by the fraction of the c.m. energy transformed into observed 
particles, 
so that, after the energy shift in $\spp$ is applied, the pp/\app\ data
reproduce reasonably well the \ep\ multiplicity data. For the mid-rapidity
density, the subsequent system development has to be taken into
account and is shown well 
  described by
 the Landau hydrodynamics
picture providing the c.m. energy is scaled according to the contribution 
of the 
participants \ct{my,ee3pp-landau}.
 \bigskip

  {\bf 7. }  
  In summary, we analyse the average multiplicity and
midrapidity density data in  pp/\app\  and in central nuclear 
interactions 
as a function of the c.m. energy per nucleon over the whole available 
range of 
the 
interaction c.m. energies, including the highest energy LHC data $\spp 
=2.36$~TeV in pp/\app\ 
collisions, 
and the  highest energy RHIC data at $\snn=200$~GeV.
 Within the framework of constituent quarks, we develop a  model which
interrelates these two variables measured in the two types of 
interactions, 
 assuming one quark of 
 each nucleon participates in pp/\app\ collision
while all three quarks (\ie\ a complete nucleon)  participate in a  
head-on heavy 
ion collision. 
 We consider these participants to form the initial zone of a collision 
which 
then develops in hydrodynamic framework, the Landau
relativistic hydrodynamic model in our case. In this approach soft 
hadron production is determined at the very early stage of the collision.
  After appropriately taking into account the contributions of the 
participants, which requires an energy-scaling factor of 1/3 in pp/\app\ 
measurements, the average multiplicity and mid-rapidity density in 
nucleon-nucleon and nucleus-nucleus interactions are found to have a 
similar c.m. energy dependence.
 The midrapidity density is found to obey a linear-log fit 
on $\snn$, while the multiplicity data follows a second-order 
log-polynomial increase  with $\snn$. A clear preference of the data on 
the 
multiplicity 
to follow the $\log^2(s)$ behaviour is observed compared to the power-law, 
the two dependencies being indistinguishable up to $\snn$ of about 300 
GeV, 
or 
$\spp \sim 1$~TeV. 
Assuming no changes in the 
multihadron production processes
with increasing energy $\spp$ of the LHC to 7 TeV, 10 TeV and 14 TeV, 
and 
looking forward to the heavy-ion data at the corresponding $\snn$  
of 2.76 TeV,
3.94 TeV and 
5.52~TeV, 
we estimate the 
multiplicities and midrapidity densities for the forthcoming data, 
using the obtained energy dependencies.
 % \bigskip
 %\medskip

\begin{acknowledgement}
We are grateful to David Plane for his help during preparation of the 
manuscript. 
\end{acknowledgement}

\end{document}